\documentclass[12pt]{article}
\usepackage{amsmath,amssymb,color}
\usepackage{cite,epsf}

\makeatletter \@addtoreset{equation}{section} \makeatother

\newcommand{\be}{\begin{equation}}
\newcommand{\ee}{\end{equation}}
\newcommand{\bea}{\begin{eqnarray}}
\newcommand{\eea}{\end{eqnarray}}

\def\fft#1#2{\frac{#1}{#2}}

\begin{document}

\begin{flushright}
March\  2008
\end{flushright}

\vspace{10pt}

\begin{center}

{\Large\bf Drag force at finite 't Hooft coupling\\ from AdS/CFT}

\vspace{30pt}

{\large Justin F. V\'azquez-Poritz}

\vspace{20pt}

{\it Department of Natural Sciences\\ Baruch College, The City University of New York, New York NY 10010} 

\vspace{10pt}

{\tt Justin\_Vazquez-Poritz@baruch.cuny.edu}

\vspace{20pt}

\centerline{{\bf{Abstract}}}
\end{center}

\noindent 
We find that the drag force for a heavy quark moving through ${\cal N}=4$ $SU(N)$ supersymmetric Yang-Mills plasma is generally enhanced by the leading correction due to finite 't Hooft coupling. For a bottom quark, the drag force increases by about $50\%$, whereas for a charm quark it increases by about $5\%$. We also discuss the drag force for the case of Gauss-Bonnet gravity.

\newpage

\section{Introduction and Summary}

According to the AdS/CFT correspondence \cite{agmoo9905}, the dynamics of open strings on a five-dimensional AdS black hole background are related to that of partons in the large $N$ and large 't Hooft coupling limit of four-dimensional ${\cal N}=4$ $SU(N)$ super Yang-Mills theory at finite temperature. 
There have been a number of proposals for using this framework to calculate measures of the rate at which partons lose energy to the surrounding plasma, such as the ultra-relativistic jet quenching parameter\footnote{The calculation of the jet quenching parameter $\hat q$ in \cite{lrw0605,lrw0612} involves a lightlike Wilson loop. However, unless there is a compelling reason for discarding the leading saddlepoint contribution to the Wilson loop, following this definition leads to zero jet-quenching parameter \cite{argyres1,argyres2}.} \cite{lrw0605} and the drag and stochastic forces of heavy quarks \cite{hkkky0605,gub0605,st0605}.

These computations are strictly valid when the 't Hooft coupling constant $\lambda\rightarrow\infty$. However, the standard values of $\alpha_{\rm SYM}=1/2$ and $N_c=3$ correspond to $\lambda=4\pi N_c\ \alpha_{\rm SYM}=6\pi$. Thus, an understanding of how these computations are affected by finite $\lambda$ corrections, which correspond to string $\alpha^{\prime}$ corrections in the dual gravity background, may be essential for more precise theoretical predictions. The first correction in the inverse 't Hooft coupling to the jet quenching parameter was found in \cite{edelstein}. Here we will extend the approach of \cite{hkkky0605,gub0605} to include the first $1/\lambda$ correction.

We find that the drag force is generally enhanced by this correction, especially for slowly moving quarks. In particular, if we take $v=0.2$ and $\alpha_{\rm SYM}$ to be $0.25$ or $0.3$ for a bottom quark, then the drag force increases from its value at infinite coupling by a factor of $1.62$ or $1.34$, respectively. The factor by which the drag force is enhanced is considerably less dramatic for lighter quarks. For instance, taking $v=0.5$ and $\alpha_{\rm SYM}$ to be $0.4$ or $0.5$ for a charm quark gives a factor of $1.06$ or $1.04$, respectively. This indicates that the corrections due to finite 't Hooft coupling are important for precise predictions of the energy loss of bottom quarks but not for lighter quarks.

At infinite coupling, the quark energy loss calculated in \cite{lrw0605} is about $20\%$ 
greater than what is calculated in \cite{hkkky0605,gub0605}. Curiously enough, the values of these two predictions for energy loss converge at $\alpha_{\rm SYM}\approx 0.3$. However, this comparison must be taken with a grain of salt, since it relies on two assumptions. Firstly, the calculation of the jet quenching parameter in \cite{lrw0605} hinges on the currently unproven claim that the leading saddlepoint contribution to the Wilson loop can be discarded. Secondly, even though \cite{hkkky0605,gub0605} uses a framework in which the quarks must be heavy, we have extrapolated these results for $v\rightarrow 1$.

We also compute the drag force for a black brane background in Gauss-Bonnet theory. We find that the drag force monotonically increases for positive Gauss-Bonnet parameter $\lambda_{GB}$ and decreass for negative $\lambda_{GB}$.

\vspace{10pt}
While this paper was in the final stages of preparation, it came to our attention that the drag force calculation for a black brane background in Gauss-Bonnet theory, as well as generic $R^2$ corrections,  has simultaneously been done in \cite{kazem}.

\section{Drag force at finite 't Hooft coupling}

Corrections in inverse 't Hooft coupling $1/\lambda$ correspond to $\alpha^{\prime}$ corrections on the string theory side. The $\alpha^{\prime}$-corrected near extremal D3-brane has the metric \cite{alpha1,alpha2}
\be\label{metricform}
ds_{10}^2=-g_{tt}\ dt^2+g_{xx}\ \delta_{ij}dx_idx^j+g_{uu}\ du^2+G_{Mn}\ dy^M dy^n\,,
\ee
where $x^M=(t,x^i,z; y^n)$, $i,j=1,2,3$ and $n=1,\dots ,5$. The metric functions are given by
\bea
g_{tt} &=& -L^2 u^2 (1-z^{-4})(1+b\ T(z)+\dots )\,,\nonumber\\
g_{xx} &=& L^2 u^2 (1+b\ X(z)+\dots )\,,\nonumber\\
g_{uu} &=& L^2 u^{-2}(1-z^{-4})^{-1} (1+b\ R(z)+\dots )\,,
\eea
where
\bea
T(z) &=& -75z^{-4}-\fft{1225}{16}z^{-8}+\fft{695}{16}z^{-12}\,,\nonumber\\
X(z) &=& -\fft{25}{16}z^{-8}(1+z^{-4})\,,\nonumber\\
R(z) &=& 75z^{-4}+\fft{1175}{16}z^{-8}-\fft{4585}{16}z^{-12}\,.\label{alphametric2}
\eea
As a matter of convenience, we express the metric components in terms of the dimensionless coordinate $z\equiv u/u_h$. There is an event horizon at $u=u_h$ and the geometry is asymptotically AdS at large $u$ with a radius of curvature $L$.

The expansion parameter $b$ can be expressed in terms of the inverse 't Hooft coupling as
\be\label{blambda}
b=\fft{\zeta(3)}{8}\lambda^{-3/2}\sim .15\lambda^{-3/2}\,.
\ee

The classical dynamics of a string in this background are described by the Nambu-Goto action
\be\label{action}
S=-\fft{1}{2\pi\alpha^{\prime}}\int d\sigma\ d\tau\ \sqrt{-G}\,,
\ee
with 
\be
G={\rm det}[g_{\mu\nu}(\partial X^{\mu}/\partial\xi^{\alpha})(\partial X^{\nu}/\partial \xi^{\beta})]\,,
\ee
where $\xi^{\alpha}=\{ \tau,\sigma\}$ and $X^{\mu}=\{ t,x_1,x_2,x_3,u\}$. The equation of motion for a string extending in one $x_i$ direction was derived in \cite{herzog} in the static gauge. For stationary solutions with the following embedding: 
\be\label{embedding}
t=\tau\,,\qquad u=\sigma\,,\qquad x_1=x(\sigma)+v\tau\,,\qquad x_2=x_3=0\,, 
\ee

\begin{figure}[h]
   \epsfxsize=3.5in \centerline{\epsffile{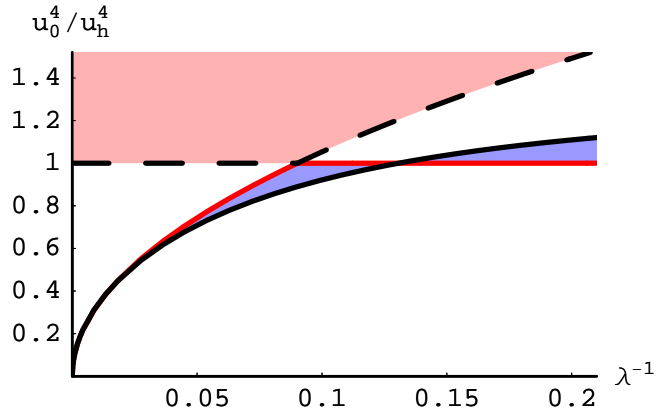}}
   \caption[FIG. \arabic{figure}.]{\footnotesize{The two real roots of $G$ lie within the shaded regions as functions of $\lambda$ and $v$.}}
\label{fig1}
\end{figure}

\begin{figure}[h]
   \epsfxsize=3.0in \centerline{\epsffile{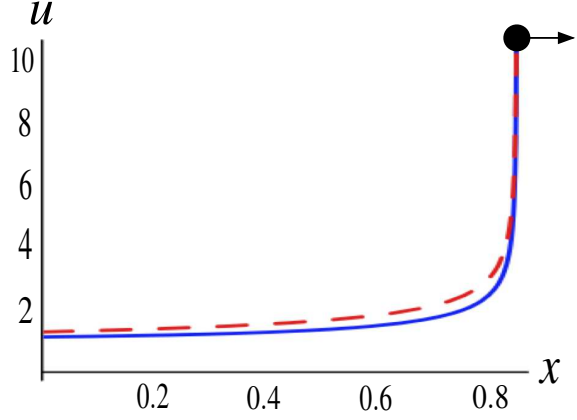}}
   \caption[FIG. \arabic{figure}.]{\footnotesize{For $v=0.2$, the string has more drag for $\lambda^{-1}=0.088$ (dashed red line) than for infinite $\lambda$ (solid blue line).}}
\end{figure}

\begin{figure}[h]
   \epsfxsize=3.5in \centerline{\epsffile{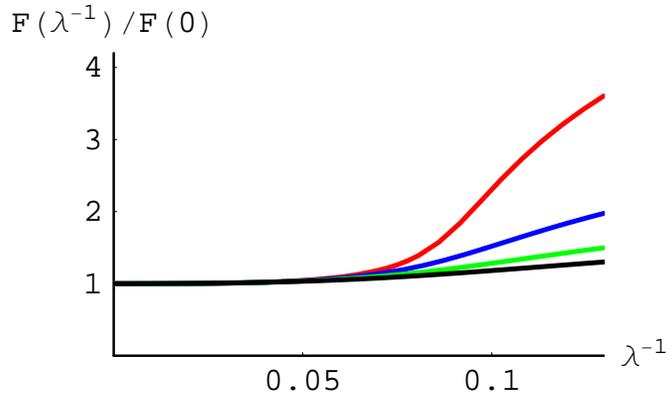}}
   \caption[FIG. \arabic{figure}.]{\footnotesize{$F(\lambda^{-1})/F(0)$ versus $\lambda^{-1}$ for $v=0.1$ (red), $0.2$ (blue), $0.3$ (green) and $0.4$ (black). The drag force is all the more enhanced for slowly moving quarks.}}
\end{figure}

\begin{figure}[h]
\begin{center}
$\begin{array}{c@{\hspace{.20in}}c}
\epsfxsize=2.58in \epsffile{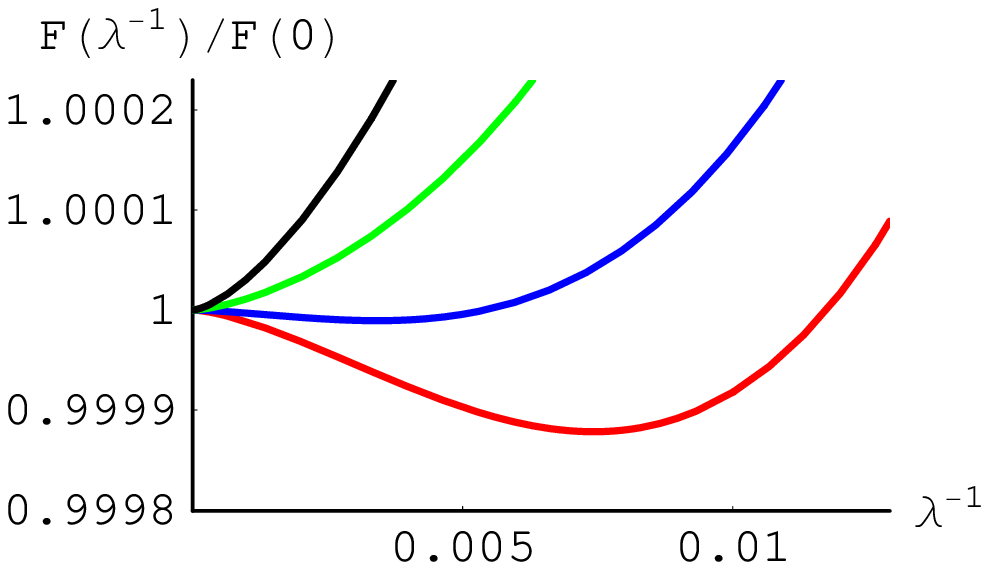} &
\epsfxsize=2.58in \epsffile{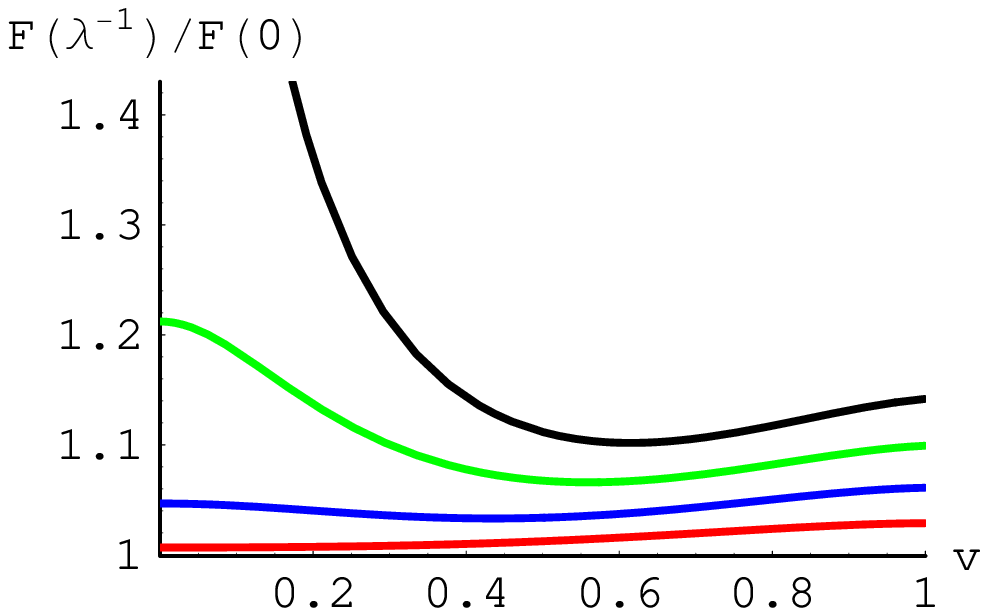}
\end{array}$
\end{center}
\caption[FIG. \arabic{figure}.]{\footnotesize{The left figure shows $F(\lambda^{-1})/F(0)$ versus $\lambda^{-1}$ blown up in the region $\lambda^{-1}<0.012$, for $v=0.1$ (red), $0.2$ (blue), $0.3$ (green) and $0.4$ (black). The figure on the right shows $F(\lambda^{-1})/F(0)$ as a function of $v$ for $\lambda^{-1}=0.03$ (red), $0.05$ (blue), $0.07$ (green) and $0.09$ (black).}}
\end{figure}
\begin{figure}[h]
   \epsfxsize=3.2in \centerline{\epsffile{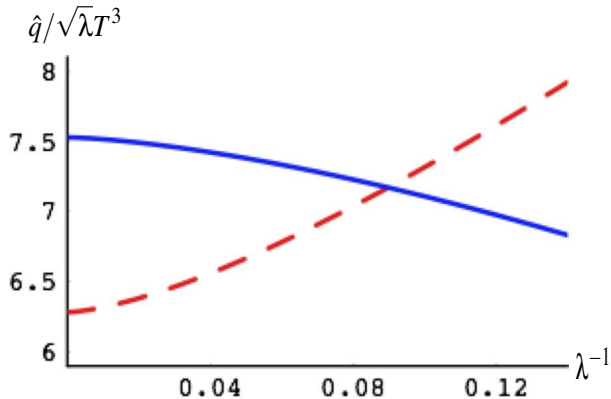}}
   \caption[FIG. \arabic{figure}.]{\footnotesize{The jet quenching parameter $\hat q$ as extrapolated from the drag force (dashed red line) agrees with the $\hat q$ calculated in \cite{edelstein} (solid blue line) at $\lambda^{-1}\approx 0.09$, which corresponds to $\alpha_{\rm SYM}\approx 0.3$.}}
\end{figure}

the equation of motion reduces to
\be
\partial_u \Big( \fft{hf_{TX}\ x^{\prime}}{\sqrt{-G}}\Big) =0\,,
\ee
where
\be
-\fft{G}{L^4}=f_{RT}-v^2 u^4 h^{-1}f_{RX}+hf_{TX}x^{\prime 2}\,.
\ee
We have defined the functions
\be
f_{XY}\equiv 1+b(X+Y)\,,\qquad h\equiv u^4-u_h^4\,,
\ee
and $x^{\prime}\equiv\partial_u x$. Solving for $x^{\prime}$ yields
\be
x^{\prime 2}=\fft{C^2 v^2(hf_{RT}-v^2u^4f_{RX})}{h^2f_{TX}(hf_{TX}-C^2v^2)}\,,
\ee
where $C$ is an integration constant.

We find that
\be
-\fft{G}{L^4}=\fft{hf_{RT}-v^2 u^4f_{RX}}{h-C^2 v^2f_{TX}^{-1}}\,,
\ee
up to first order in $b$ in both the numerator and the denominator. In order to avoid parts of the string moving faster than the local speed of light, $-G$ must be positive definite all along the string. Thus, any real roots of the numerator of $G$ which lie outside of the event horizon must also be roots of the denominator. The numerator $hf_{RT}-v^2 u^4f_{RX}$ is a fourth-order polynomial expression in $u^4$, which has two complex roots and two real roots. 

These two real roots as a function of $\lambda^{-1}$ and for various values of $v$ are shown in Figure 1. 
The larger root lies within the pink region. This root approaches the dashed line as $v\rightarrow 0$ and goes to infinity as $v\rightarrow 1$. The smaller root lies within the violet region which is enclosed by the solid red line, corresponding to $v\rightarrow 0$, and the solid black line, which corresponds to $v\rightarrow 1$. Note that the larger root always lies outside of the event horizon for nonvanishing velocity. On the other hand, the smaller root lies within the horizon for $\lambda^{-1}<0.131$ and outside of the horizon for $\lambda^{-1}>0.131$. This could imply that certain restrictions must be imposed on the mass or velocity of the quark for $\lambda^{-1}>0.131$. On the other hand, the appearance of multiple roots outside of the horizon might simply be an indication that higher-order corrections in inverse 't Hooft coupling must be considered for $\lambda^{-1}> .131$. Thus, we restrict ourselves to $\lambda^{-1}<0.131$.

Demanding that the numerator and denominator of $G$ vanish at the same radius outside of the horizon enables us to solve for $C$ as a function of $v$ and $\lambda$. $C$ can then be plugged back into the equation of motion, which can be numerically integrated in order to plot the string configuration. As an example, we have plotted the result for $v=0.2$ in Figure 2, from which it can be seen that the string has more drag for $\lambda^{-1}=0.088$ ($\alpha_{\rm SYM}=0.25$) than for infinite $\lambda$.

It has been found that energy and momentum flow from one end of this string to the other at a rate
\bea
\fft{dE}{dt} &=& \fft{1}{2\pi\alpha^{\prime}} Cv\,,\nonumber\\
\fft{dP}{dt} &=& \fft{1}{2\pi\alpha^{\prime}} C\,.
\eea
Thus, the drag force $F$ is proportional to $C$. We will consider the ratio $F(\lambda^{-1})/F(0)$, where $F(\lambda^{-1})$ is the drag force for nonvanishing $\lambda^{-1}$, and $F(0)$ is the drag force for zero $\lambda^{-1}$.

$F(\lambda^{-1})/F(0)$ as a function of $\lambda^{-1}$ is shown in Figure 3, from which it can be seen that the quark energy loss is enhanced by the corrections due to finite 't Hooft coupling. This enhancement is all the more dramatic for slowly moving quarks at smaller 't Hooft coupling, such as bottom and charm quarks. Recalling that $\lambda=4\pi N_c\ \alpha_{\rm SYM}$ and setting $N_c=3$, we can consider a sampling of values for $\alpha_{\rm SYM}$, $v$ and $F(\lambda^{-1})/F(0)$. In particular, we find that the drag force for a bottom quark (we set $v=0.2$ and $\alpha_{\rm SYM}=0.25$ or $0.3$) would be enhanced by the $1/\lambda$ correction by a factor of $1.62$ or $1.34$, respectively. On the other hand, the drag force for a charm quark (we set $v=0.5$ and $\alpha_{\rm SYM}=0.4$ or $0.5$) is enhanced by a factor of $1.06$ or $1.04$, respectively.

The right plot in Figure 4 shows that the drag force is slightly decreased by the inverse 't Hooft coupling correction for small $v$ and small $\lambda^{-1}$. This can also be seen by expanding $F(\lambda^{-1})/F(0)$ in both $v$ and $\lambda^{-1}$ as
\be
\fft{F(\lambda^{-1})}{F(0)}\approx (1-.5 \lambda^{-3/2}+298.6\lambda^{-3}+\dots ) +7.9\lambda^{-3/2}( 1-1.4\lambda^{-3/2}+\dots ) v^2+\dots\,.
\ee
For small $v$, finite $\lambda$ has the effect of decreasing the drag force for $\lambda^{-1}\le 0.014$, and enhancing the drag for larger $\lambda^{-1}$. However, according to the right plot in Figure 4, the decrease in drag is on the order of $0.01\%$, and might therefore merely be a remnant of not including the higher-order $1/\lambda$ corrections.

The left plot in Figure 4 shows that $\fft{F(\lambda^{-1})}{F(0)}$ is not a monotonic function of $v$.
However, we should keep in mind that the drag force is computed within a non-relativistic framework which applies for heavy quarks. Thus, our results for larger velocities must be taken with a grain of salt.

Nevertheless, one can attempt to extrapolate the ultra-relativistic jet quenching parameter from the drag force. In the $\lambda\rightarrow\infty$ limit, it was found that $\hat q/\sqrt{\lambda}T^3\approx 6.28$ as $v\rightarrow 1$ \cite{hkkky0605}. As shown in Figure 5, this quantity increases (dashed red line) with $\lambda^{-1}$. If it turns out that the leading saddlepoint contribution to the Wilson loop can be discarded then the non-perturbative definition of the jet quenching parameter given in \cite{lrw0605} provides an independent method for calculating $\hat q$. In the limit $\lambda\rightarrow\infty$, it was found that $\hat q/\sqrt{\lambda}T^3\approx 7.53$ \cite{lrw0605}. The $1/\lambda$ correction for this $\hat q$ was considered in \cite{edelstein}, and is also shown in Figure 5 (solid blue line). Note that these two curves intersect at $\lambda^{-1}\approx 0.09$, which corresponds to $\alpha_{\rm SYM}\approx 0.3$.

\section{Drag force in Gauss-Bonnet gravity}

We will now consider how curvature-squared terms affect the drag force, for the particular case of Gauss-Bonnet gravity. It has recently been shown that the conjectured lower bound of $1/4\pi$ on the viscosity-to-entropy ratio \cite{bound1,bound2} can be violated by curvature-squared terms \cite{R21,R22}, which makes it especially interesting to consider them. Gauss-Bonnet gravity has a number of useful properties that are not shared with theories involving more general curvature-squared terms. For example, the action involves only second derivatives of the metric \cite{zwiebach}. Also, exact solutions have been obtained in \cite{gb1,gb2}.

Gauss-Bonnet gravity is defined by the action \cite{zwiebach}
\be
S=\fft{1}{16\pi G_N} \int d^5x \sqrt{-g}\Big[ R+\fft{12}{L^2}+\fft{\lambda_{GB}}{2} L^2 (R^2-4R_{\mu\nu}R^{\mu\nu}+R_{\mu\nu\rho\sigma}R^{\mu\nu\rho\sigma})\Big]\,.
\ee
There is an exact black brane solution \cite{gb1,gb2} whose metric is of the form (\ref{metricform}) with
\be
g_{tt}=-L^2 u^2fk\,,\qquad g_{xx}=L^2u^2\,,\qquad g_{uu}=\fft{L^2}{u^2f}\,,
\ee
and
\be
f=\fft{1}{2\lambda_{GB}}\Big[ 1-\sqrt{1-4\lambda_{GB}\Big(1-\fft{u_h^4}{u^4}\Big) }\Big]\,,\qquad k=\fft12 (1+\sqrt{1-4\lambda_{GB}})\,.
\ee
The scaling factor $k$ for time ensures that the speed of light in the boundary theory is unity. We assume that $\lambda_{GB}\le 1/4$, since beyond this point there is no vacuum AdS solution.

For stationary string solutions of the Nambu-Goto action (\ref{action}) with the embedding given by (\ref{embedding}), the equation of motion reduces to
\be
\partial_u \Big( \fft{fu^4x^{\prime}}{\sqrt{-G}}\Big) =0\,,
\ee
where
\be
-\fft{G}{L^4}=k-v^2f^{-1}+kfu^4 x^{\prime 2}\,.
\ee
Solving for $x^{\prime}$ yields
\be
x^{\prime 2}=\fft{C^2 v^2(kf-v^2)}{kf^2u^4(kfu^4-C^2v^2)}\,,
\ee
where $C$ is an integration constant.

We find that
\be
-\fft{G}{L^4}=\fft{ku^4(kf-v^2)}{kfu^4-C^2v^2}\,.
\ee
In order for $-G$ to remain positive everywhere on a string that stretches from the boundary to the horizon, we require that
\be
C^2=\Big(1-\fft{v^2}{k}+\fft{\lambda_{GB}v^4}{k^2}\Big)^{-1} u_h^4\,.
\ee
This yields
\be
\fft{F(\lambda_{GB})}{F(0)}=\Big( \fft{1+\sqrt{1-4\lambda_{GB}}-2\lambda_{GB}(1+v^2)}{2}\Big)^{-1/2}\,.
\ee
This ratio of drag forces is plotted in Figure 6 for various values of the velocity, from which we see that the drag force is generically increased for positive $\lambda_{GB}$ and decreased for negative $\lambda_{GB}$. Also, the effect of nonzero $\lambda_{GB}$ is slightly enhanced for quarks moving close to the speed of light. 

\begin{figure}[h]
   \epsfxsize=3.5in \centerline{\epsffile{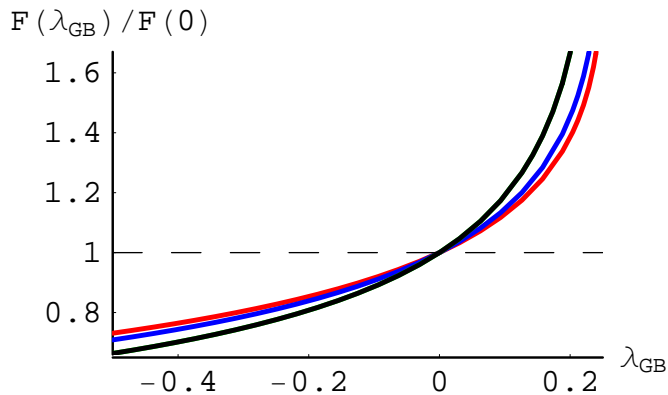}}
   \caption[FIG. \arabic{figure}.]{\footnotesize{$F(\lambda_{GB})/F(0)$ as a function of $\lambda_{GB}$ for $v=0.1$ (red), $0.5$ (blue) and $0.9$ (black) for the Gauss-Bonnet black brane background. Depending on the sign of $\lambda_{GB}$, the drag force is either increased or decreased in a monotonic manner.}}
\end{figure}

\begin{figure}[h]
   \epsfxsize=3.2in \centerline{\epsffile{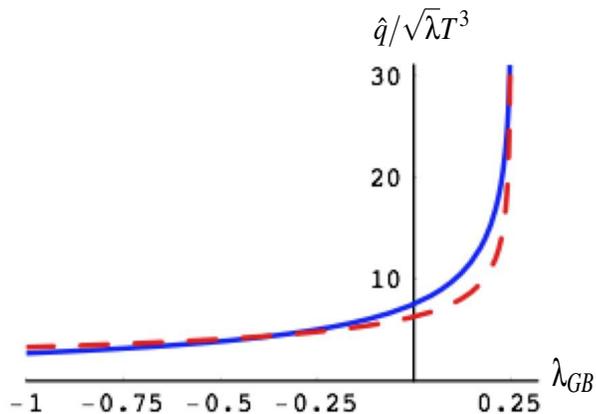}}
   \caption[FIG. \arabic{figure}.]{\footnotesize{The jet quenching parameter $\hat q$ as extrapolated from the drag force (dashed red line) agrees with the one given by the non-perturbative definition in \cite{lrw0605} (solid blue line) at both $\lambda_{GB}\approx -0.354$ and $0.249$.}}
\end{figure}

As a curiosity, we compare the ultra-relativistic jet quenching parameter $\hat q$ as extrapolated from the above drag force with the $\hat q$ resulting from the non-perturbative definition given in \cite{lrw0605}. As shown in Figure 7, these two results for $\hat q$ agree at $\lambda_{GB}\approx -0.354$ and $0.249$. However, as previously mentioned, such a comparison can only be made provided that the leading saddlepoint contribution to the Wilson loop can be discarded in the calculation of \cite{lrw0605} and that the energy loss results of \cite{hkkky0605,gub0605} are not drastically affected by relativistic effects.

\section*{Acknowledgments}

I would like to thank Philip Argyres, Kazem Bitaghsir, Mohammad Edalati and Jamal Jalilian-Marian for helpful discussions.

\end{document}